\documentclass[prb,preprint,preprintnumbers,amsmath,amssymb,unsortedaddress]{revtex4}

\usepackage{graphicx}
\usepackage{dcolumn}
\usepackage{bm}

\usepackage{color}

\begin{document}
 

\title{Current sheets at three-dimensional magnetic nulls: Effect of compressibility}

\author{D.~I.~Pontin} 
\email[]{dpontin@maths.dundee.ac.uk}
\altaffiliation[Now at: ]{Division of Mathematics, University of Dundee, Dundee, Scotland}
\author{A.~Bhattacharjee}
\affiliation{Space Science Center and Center for Magnetic Self-Organization, University of New Hampshire, Durham, New Hampshire, USA}
\author{K.~Galsgaard}
\affiliation{Niels Bohr Institute, University of Copenhagen, Copenhagen,
  Denmark}

\date{\today}

\begin{abstract}
The nature of current sheet formation in the vicinity of three-dimensional (3D)
magnetic null points is investigated. The particular focus is upon the
effect of the compressibility of the plasma on the qualitative and
quantitative properties of the current sheet. An initially potential 3D null is subjected to shearing perturbations, as in a previous paper [Pontin {\it et al.}, Phys. Plasmas, in press (2007)]. 
It is found that as the incompressible limit is approached, the collapse of the null point is
suppressed, and an approximately planar current sheet aligned to the
fan plane is present instead. This is the case regardless of whether the spine or fan of the null is sheared. 
Both the peak current and peak reconnection rate are reduced. The results have a bearing on previous analytical solutions for steady-state reconnection in incompressible plasmas,  implying that fan current sheet solutions are dynamically accessible, while spine current sheet solutions are not.
\end{abstract}

\pacs{Valid PACS appear here}

\maketitle

\section{Introduction}
In astrophysical plasmas, such as the solar corona,
the three-dimensional (3D) magnetic field topology is often highly complex. In such
complex 3D magnetic fields, where traditional two-dimensional (2D) X-point magnetic reconnection
models may no longer be applicable, determining the sites at which dynamic phenomena
and energy release may occur is a crucial and non-trivial problem. Due to the
typically very high Lundquist number, such events occur only at locations where
intense currents (singular under an ideal MHD evolution) may form. One
such site is a 3D magnetic null point
(e.g.~Refs.~[\onlinecite{klapper1996,priest1996,bulanov1997,pontincraig2005,pontinbhat2007}]). 
The nature of current sheet formation  at such 3D nulls is investigated here.

3D null points are predicted to be present in abundance 
in the solar corona (e.g.~Refs.~[\onlinecite{longcope2003,close2004}]). 
Furthermore, there is observational evidence that reconnection at a 3D null 
may be important in some solar flares \citep{fletcher2001}, as well as in
eruptive phenomena in active regions \cite{ugarteurra2007}. 
In addition, the first in situ observation \citep{xiao2006} of reconnection occurring at a 3D
null point in the Earth's magnetotail has recently been made by
the Cluster spacecraft.
Moreover, current growth at 3D nulls has been observed in the laboratory
\cite{bogdanov1994}. 

The magnetic field topology and geometry in the vicinity of such a null can be
described by the two sets of field lines which asymptotically approach, or
recede from, the null. A pair of field lines approach (recede from) the null
from opposite directions, defining the `spine' (or $\gamma$-line) of the
null. In addition, an
infinite family of field lines recede from (approach) the null in a surface
known as the fan (or $\Sigma$-) plane 
(see Refs.~[\onlinecite{parnell1996,priest1996}]).

To this point, many studies of the MHD behaviour of 3D nulls have
been kinematic, see
e.g.~Refs.~[\onlinecite{lau1990,priest1996,pontin2004,pontinhornig2005}]. However, 
a few solutions to the full set of MHD equations do exist for reconnection at
current sheets located at 3D nulls, in incompressible plasmas. These
incompressible solutions are based upon the technique first proposed by
Craig \& Henton \cite{craig1995} for the 2D reconnection problem. The
solutions describe steady-state current sheets aligned to the fan and spine
of the null \cite{craig1996}. Time-dependent solutions for the fan current
sheets also exist \cite{craig1998}.

In a previous paper---Ref.~[\onlinecite{pontinbhat2007}], hereafter referred to as paper
I---we investigated the behaviour of 3D null points which are subjected to
shearing boundary motions, and found that current sheets formed at the
null. In this paper we consider the effect of moving from the compressible towards the
incompressible limit, which is found to have a profound effect on both the
quantitative and qualitative properties of the current sheet. This is highly
relevant when it comes to comparing the observed current sheet formation with
the analytical models, which must by necessity invoke various simplifications. Typically, the plasma in the solar atmosphere or Earth's magnetosphere is compressible. Thus it is of great interest to understand the relationship between this regime and the incompressible approximation, upon which much of the previous theory has been based. 

The remainder of the paper is set out as follows. In Sec.~\ref{idealgas} we
briefly review the previous results of paper I. In Sec.~\ref{fansheets} we
describe simulations in which we move towards the incompressible limit, and
in Sec.~\ref{analytical}, we discuss the relation of our results to
analytical incompressible solutions, and the implications for their dynamic
accessibility. In Sec.~\ref{fandrivesec} we consider the case where we drive
across the fan instead of across the spine of the null, and finally in
Sec.~\ref{conc} we present a summary.


\section{Behaviour in a compressible plasma}\label{idealgas} 
In paper I, we discussed the evolution of the magnetic field in the vicinity
of a generic 3D magnetic null. We demonstrated by means of a kinematic solution
that an evolution of the null which acts to change the angle between the spine
and fan (such that the ratios of the null eigenvalues change in time) is
prohibited in ideal MHD. We then went on to present the results of numerical
simulations, which demonstrated the formation of strong current concentrations
at the null in response to boundary perturbations. Simulation runs based on
the same numerical code are presented below
(for further details on the numerical scheme, see Refs.~[\onlinecite{galsgaard1997,archontis2004}]).

At $t=0$ the magnetic field in the domain is given by ${\bf
  B}=B_0\left(-2x,y,z\right)$, which defines a 3D null whose spine lies along
  the $x$-axis, and whose fan is in the $x=0$ plane. ${\bf J}=0$, and so
  taking the density ($\rho$) and internal energy ($e$) of the plasma to be uniform at $t=0$
  we begin with an equilibrium 
  [we take $\rho=1$, $e=\beta\gamma/(\gamma-1)$, where $\beta$ is a constant which
  determines the plasma-$\beta$ (which is of course spatially dependent) and $\gamma$ is the ratio of specific
  heats]. All of the  domain boundaries are line-tied, and are located at $[x,y,z]=[\pm X_l, \pm Y_l, \pm Z_l]$. 
  The configuration is then perturbed by imposing a plasma flow on
  the $x$-boundaries, while the $y$- and $z$-boundaries are placed
  sufficiently far away that there is insufficient time for information to
  propagate to them and back to the null before the simulations are halted.
  
A single time unit in the simulations is equivalent to the Alfv{\' e}n travel time across a unit length
in a plasma of density $\rho=1$ and uniform magnetic field of modulus 1. The resistivity is taken to be 
uniform, with its value being based upon the dimensions of the domain. Note that at $t=0$, 
${\bf B}$ is scale-free as it is linear, and thus, the actual value of $\eta$ is somewhat arbitrary until we fix a physical length scale to associate with the size of our domain.

The boundary driving takes the form on each boundary of two distorted vortices of opposite sense, which combine to provide the desired effect of advecting the spine in the ${\hat {\bf y}}$ direction, 
in opposite senses on opposite boundaries ($x=\pm X_l$) [see paper I, Eq.~(19) and Fig.~2(b)].
In the majority of the runs described, the driving profile is transient, with
its time dependence defined by
\begin{equation}\label{tdepend}
V_0(t) = v_0 \left( \left(\frac{t-\tau}{\tau} \right)^4 -1 \right)^2,\qquad
0 \leq t \leq 2\tau, 
\end{equation}
$v_0$, $\tau$ constant, so that the driving switches on at $t=0$ and off again
at $t=2\tau$.
The result is that a current concentration
forms at the null, which is expected to be singular in the ideal limit
\cite{pontincraig2005}. During the early evolution, a stagnation flow,
accelerated by the Lorentz force (but opposed by the plasma pressure) acts to
close up the spine and fan towards one another locally at the null. The initial null is unstable to such a collapse of the spine and fan in any plane containing the spine, with the $z=0$ plane being selected by the orientation of the boundary driving. It is precisely this collapse that was shown in the kinematic solution earlier in the paper to be prohibited under ideal MHD. Thus it must be facilitated by non-ideal processes.

Due to this local collapse, a current sheet forms which
typically spans the collapsed spine and fan, with a
tendency to spread along the fan surface (especially for weaker driving). 
Accompanying the current growth is the development of a component of ${\bf E}$
parallel to ${\bf B}$ ($E_{\|}$), signifying a breakdown of ideal behaviour,
and magnetic reconnection. The integral of this quantity along the magnetic
field line in the fan perpendicular to the shear plane can be shown to give a
physically meaningful measure of the reconnection rate---giving the rate of
flux transfer across the fan (separatrix) surface, see Ref.~[\onlinecite{pontinhornig2005}].

An examination of the quantitative properties of the current sheet showed that 
the peak current, peak reconnection rate, and sheet dimensions all scale
linearly with the modulus of the driving velocity. In addition, under continual boundary shearing, the current sheet appears to grow in size and modulus indefinitely (rather than being controlled by any self-regulating mechanism). This type of behaviour is also observed in 2D `forced' or `driven' reconnection simulations. The nature of the current sheet seems to be controlled at all times by the degree of boundary displacement of the spine and fan (and how quickly this displacement is attained), and so it may share some properties (at a given time) with the 2D `non-uniform reconnection' regimes (see Ref.~[\onlinecite{priest2000}] for a review). Care should be taken, however, in drawing parallels with either of these 2D models, since each involves an inflow of plasma through the boundaries. By contrast, in our simulations the driving velocity is imposed parallel to the boundaries.

In paper I we considered the case of a monatomic ideal gas, that is we took
the ratio of specific heats, $\gamma=5/3$. It is straightforward to see that
the incompressible limit may be reached formally by letting $\gamma
\rightarrow\infty$. Taking the time-derivative of the polytropic equation of
state, $p/\rho^{\gamma}=const$, and substituting for $d\rho /dt$ using the
continuity equation gives
\begin{displaymath}
\nabla\cdot{\bf v} = \frac{1}{\gamma p} \frac{d p}{d t}.
\end{displaymath}


\section{Towards incompressible limit}\label{fansheets}
We repeat here the simulations described in paper I, with increased values of
$\gamma$. This is somewhat problematic numerically (due to the increased wave
speeds in the system), but in fact it turns out that even for moderately large values of
$\gamma$, the differences are striking.

\subsection{Qualitative differences}
The parameters chosen for the simulation runs closely follow those taken in
paper I, and are as 
follows. We take $B_0=1$, the driving strength $v_0=0.01$, $\tau=1.8$,
$A_d=80$ (boundary driving localisation), $\beta=0.05$,
$\eta=5\times10^{-4}$, and the numerical domain has dimensions $X_l=0.5$, $Y_l=Z_l=3$. 
The resolution of the simulations is $128^3$, on a non-uniform mesh with smallest grid spacing near the null to achieve higher resolution there; $\delta x\sim0.0035$ and $\delta y, \delta z \sim 0.020$.

\begin{figure}
\centering
(a)\includegraphics[scale=0.4]{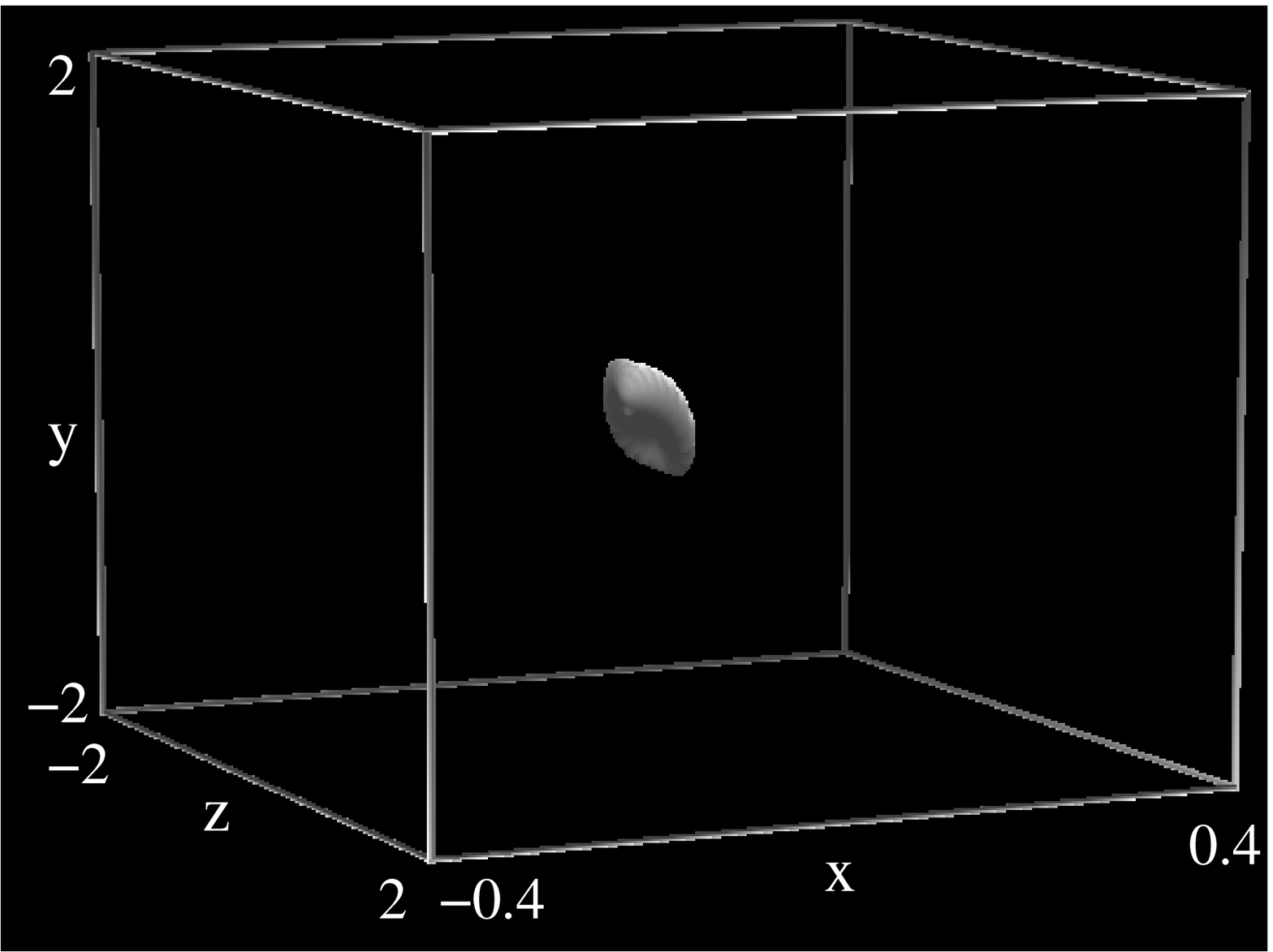}
(b)\includegraphics[scale=0.4]{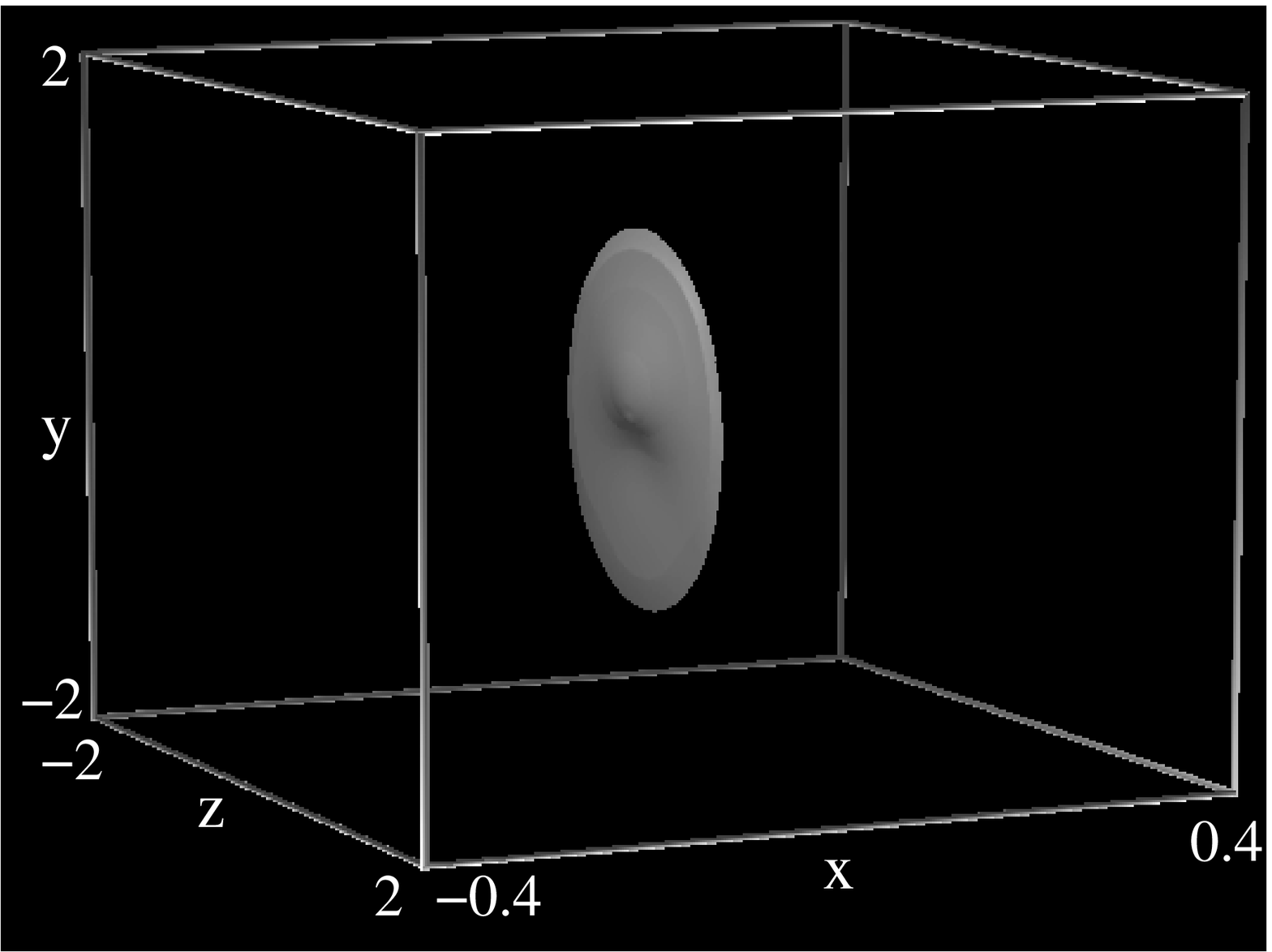}\\
(c)\includegraphics[scale=0.4]{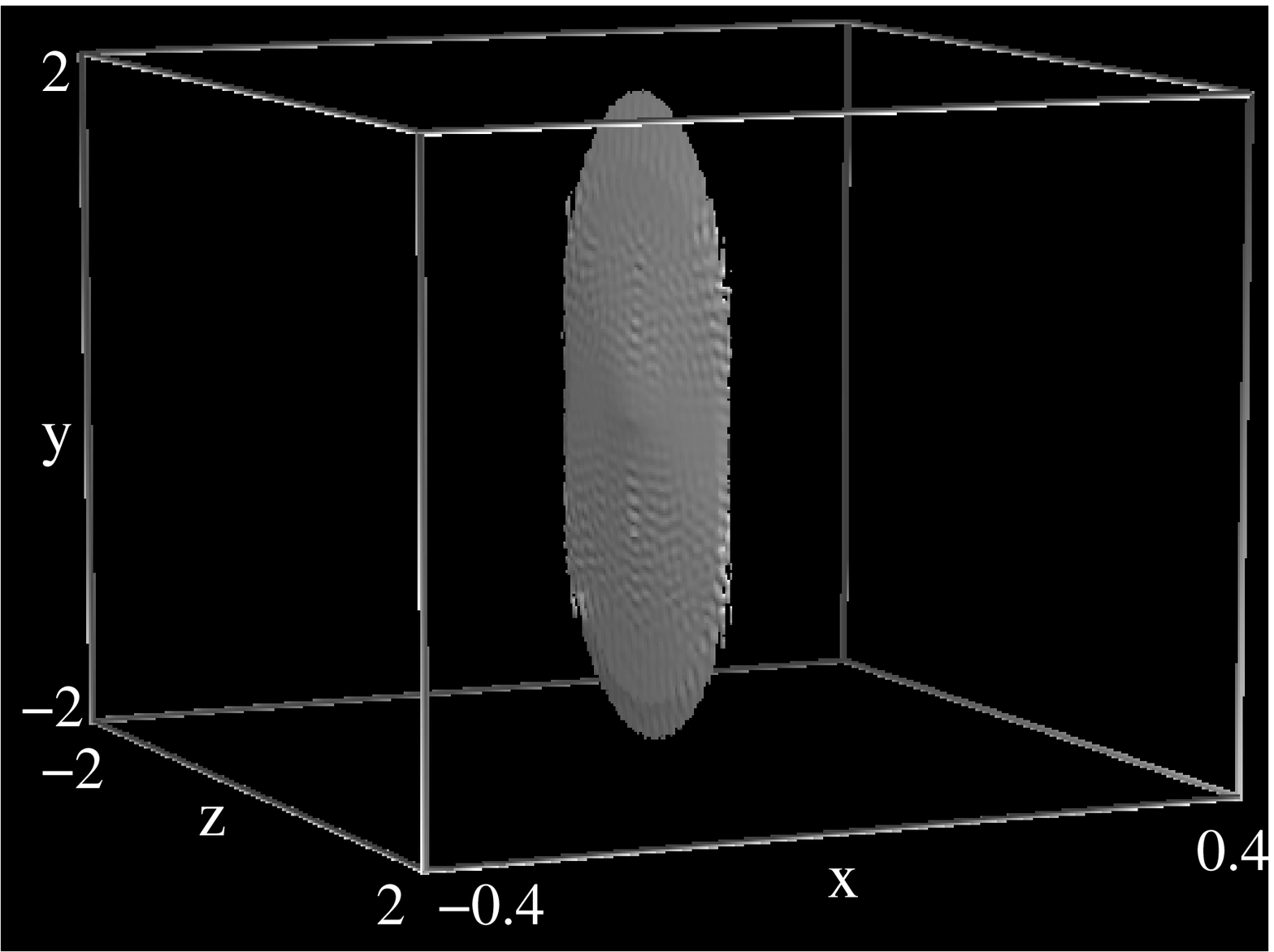}
\caption{Isosurfaces of $|{\bf J}|$ at 50$\%$ of maximum, at the time of its temporal
  peak, for (a) $\gamma=5/3$, (b) $\gamma=10/3$ and (c) $\gamma=10$.}
\label{jisos}
\end{figure}
\begin{figure}
\centering
\includegraphics[scale=0.55]{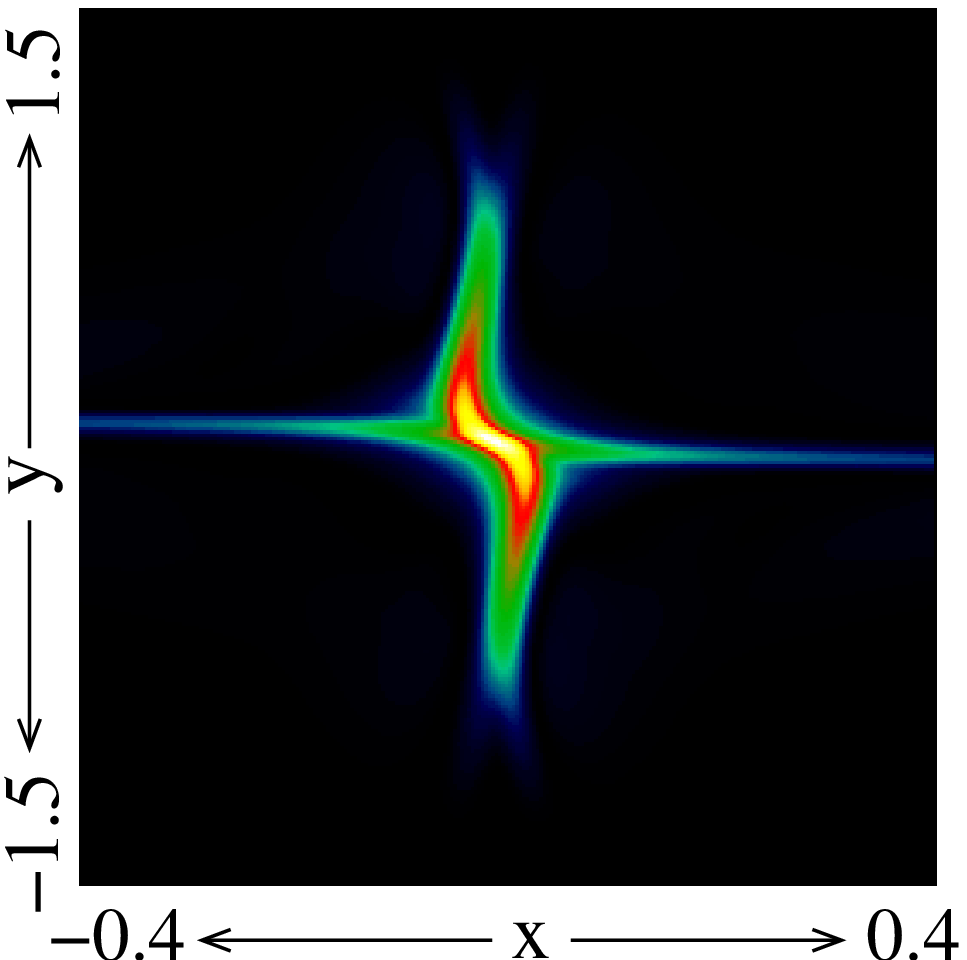}
\includegraphics[scale=0.55]{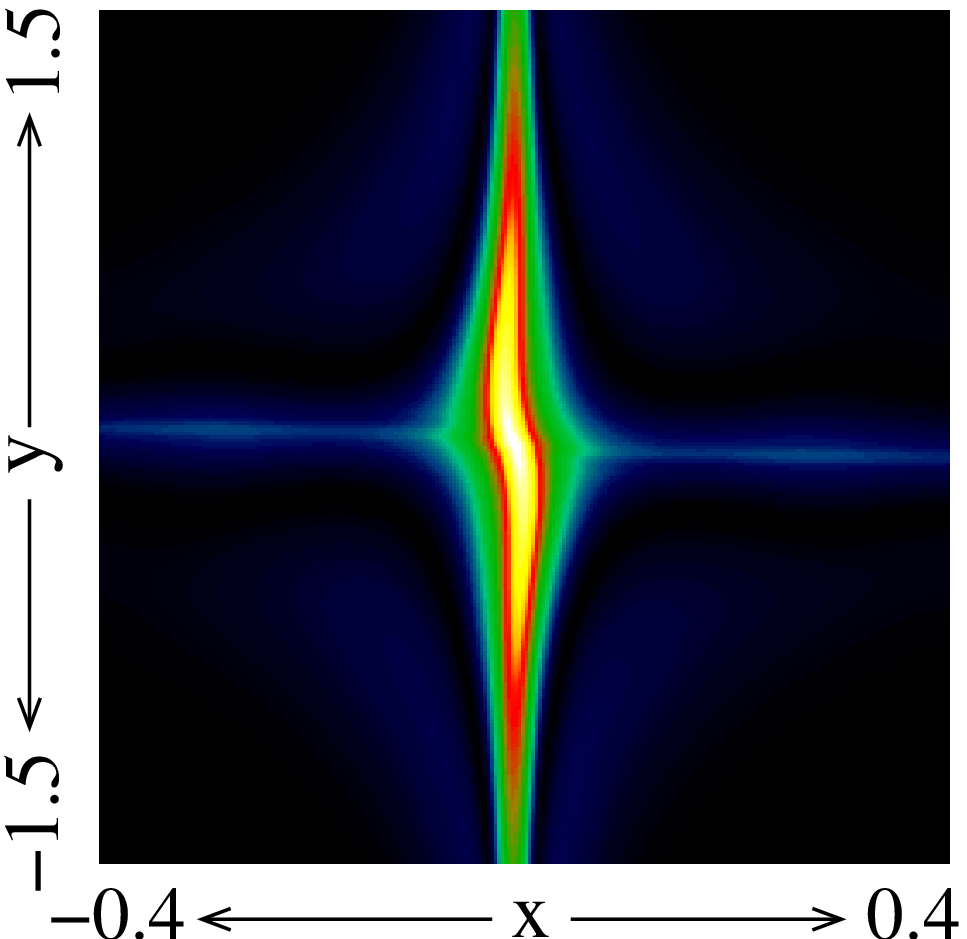}
\includegraphics[scale=0.55]{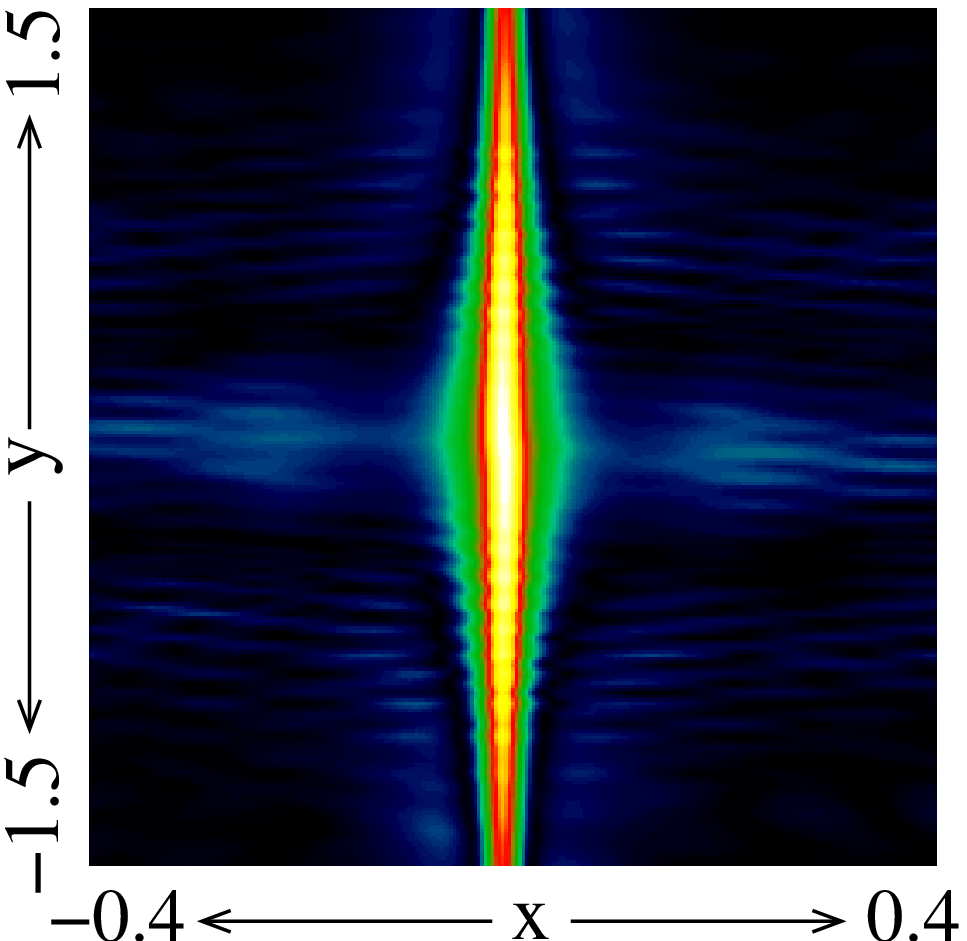}
\caption{(Colour online) Current density $|{\bf J}|$ in the $z=0$ plane, at the time of its temporal
  peak, for (a) $\gamma=5/3$, (b) $\gamma=10/3$ and (c) $\gamma=10$.}
\label{jxy}
\end{figure}
As the driving begins ($t=0$), a disturbance propagates along the spine (and
nearby field lines), and focuses at the null. For $\gamma=5/3$, the null point
`collapses' with the spine and fan closing up towards one another. A strongly
focused current sheet spans the spine and fan. However, for larger values
of $\gamma$, the
current concentration distributes itself along the fan surface, becoming more
weakly focused at the null for increasing $\gamma$ (see Figs.~\ref{jisos},
\ref{jxy}). Furthermore, the fan surface remains increasingly planar at larger
$\gamma$ (see Fig.~\ref{jxy}), and also the spine and fan do not collapse
towards each other to the same extent. This is demonstrated in
Fig.~\ref{scaling}(f), where the minimum angle between the spine
and fan ($\theta_{min}$) is plotted for runs with various values of
$\gamma$. We observe that even for 
$\gamma=10$, although the current sheet is approximately planar (at $x=0$),
the minimum angle between the spine and fan is still significantly less than
$\pi/2$. This is because the spine is still driven towards the fan by the
boundary driving (most of the stress from which is taken up in the weak field
region around the null itself), even though the fan remains approximately in the
$x=0$ plane rather than collapsing sympathetically towards the spine.

It is worth noting that the above described behaviour also depends on other
parameters in the simulation. For example, how effectively the null collapses is also dependent on the driving speed, with greater collapse and stronger
focusing of the current sheet for larger $v_0$ (see paper I). Therefore larger
values of $\gamma$ are likely to be required in order to render the fan
approximately planar for larger $v_0$, and also for larger $\tau$ (longer
driving time). The plasma-$\beta$ is also a crucial parameter, since we find
that increasing $\beta$ has a very similar effect to increasing $\gamma$. 
It is natural to expect this on physical grounds, since increasing either parameter has the effect of increasing the sound speed, and reducing the effect of magnetic forces in plasma compression.
Finally,  since the null collapse is driven by the
Lorentz force, a thinner more intense current sheet, which will form for a
lower value of $\eta$, will increase the degree of collapse. Thus, the extent to which the null collapses and the current focusses at the null is dependent on a combination of the driving velocity ($v_0$, $\tau$) and the plasma parameters $\gamma$, $\beta$ and $\eta$.

An obvious question when examining the above results is whether the planar
current sheet in the fan plane for large $\gamma$ is a result of the
symmetry of the configuration, with the null at the centre of the domain and
the fan plane parallel to the driving boundaries. We therefore re-ran the
simulations at large $\gamma$ with the null point rotated by a finite angle in
the $xy$-plane (so that the spine and fan were no longer parallel to the
boundaries). In this case, a planar current sheet still forms in the fan, and
thus our results seem general in this respect.

\begin{figure}
\centering
\includegraphics[scale=0.5]{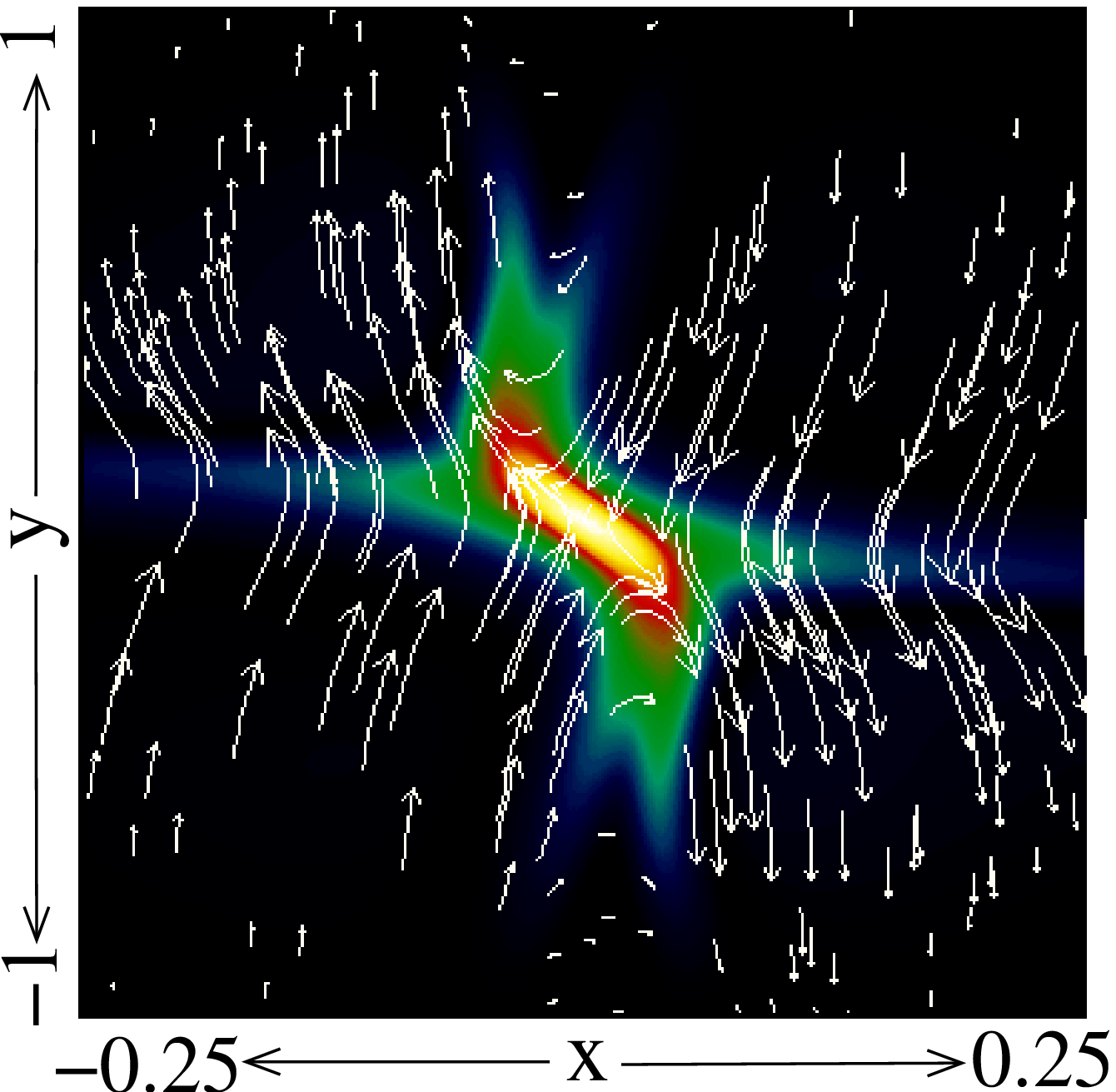}
\includegraphics[scale=0.5]{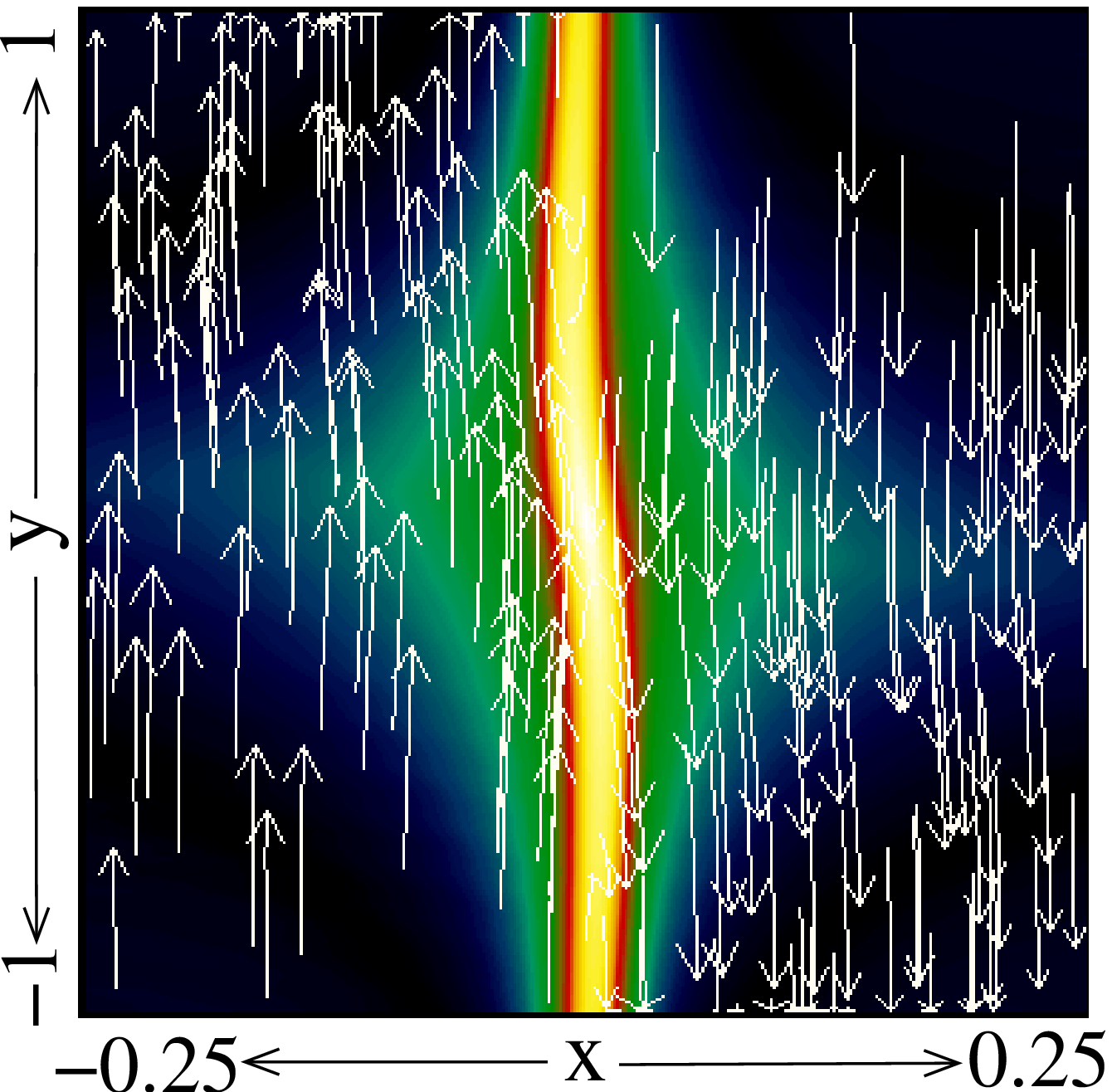}
\caption{(Colour online) Plasma flow in the $z=0$ plane at $t\approx2.5$, for (a) $\gamma=5/3$
and (b) $\gamma=10$. Background shading shows $|{\bf J}|$.}
\label{vxy}
\end{figure}

Accompanying the changing current localisation as we move towards the
incompressible limit is a change in the behaviour of the plasma flow. This
again signifies the fact that the fan of the null remains increasingly
planar. For $\gamma=5/3$, a stagnation flow is typically set up, which is
accelerated by the Lorentz force (and opposed by the plasma pressure
gradient), and which closes up the spine and fan. However, for larger $\gamma$
this flow is absent, and instead $v_x$ is approximately zero, and the flow is
roughly parallel with the driving boundaries (see Fig.~\ref{vxy}).

Finally, it should be noted that all of the above considerations are the same
as for the case of a 2D X-point. That is, repeating the above simulations but
with the magnetic field at $t=0$ defined by ${\bf B}=B_0(-x,y,0)$, we see the
same trend. For $\gamma=5/3$ the X-point collapses, forming a current sheet
which locally spans the two separatrices (a `Y-point' appearance), but for large $\gamma$ the X-point
collapse is suppressed, and the current spreads along the (unsheared)
separatrix (as in Ref.~[\onlinecite{craig1995}]).

\subsection{Quantitative differences}
\begin{figure}
\centering
\includegraphics{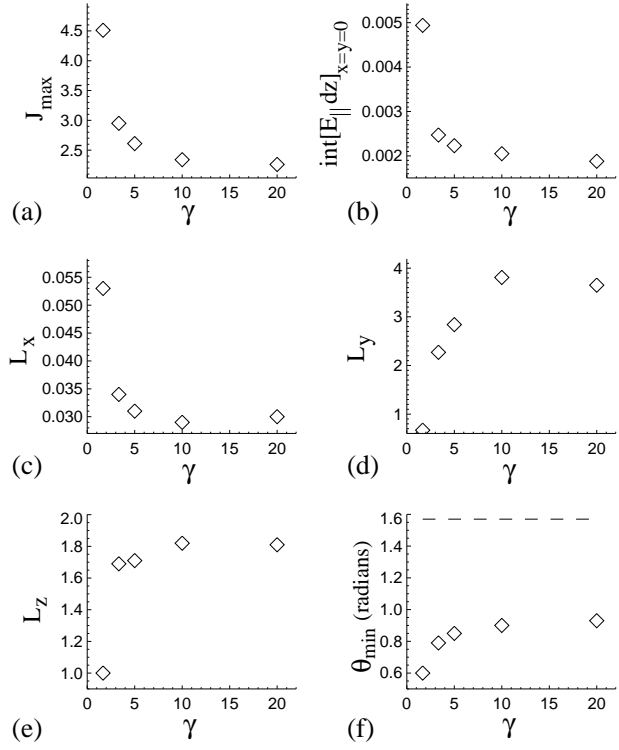}
\caption{Scaling with $\gamma$ of (a) the peak current ($J_{max}$), (b) the peak
  reconnection rate ($\int E_{\|}$), (c, d, e) the current sheet FWHM at time of peak
  current ($L_x,L_y,L_z$) and (f) the minimum angle between the spine and fan
  ($\theta_{min}$), where the dashed line indicates the value $\theta=\pi/2$.} 
\label{scaling}
\end{figure}

It is not only the qualitative properties of the current sheet which are
affected by changing the plasma compressibility. Accompanying the spreading of the current sheet
along the fan for increased $\gamma$ is a decrease in the peak current and
reconnection rate in the simulation, see Fig.~\ref{scaling}(a, b). 
The rate of change of each quantity around $\gamma=5/3$ is much greater than
that around $\gamma=10$, implying that even for this moderate value of
$\gamma$, the behaviour is already a fairly good approximation to the
incompressible limit (for all other parameters fixed). The change in geometry
of the current sheet is evidenced 
by the variation in the dimensions of the region of high $|{\bf J}|$, $L_x$, 
$L_y$ and $L_z$ (measured at the time
of current maximum, by the full-width-at-half-maximum (FWHM) in each
coordinate direction). $L_y$ and $L_z$ increase with $\gamma$, showing how
the current 
spreads along the fan surface as we move towards the incompressible limit [Fig.~\ref{scaling}(d, e)]. On
the other hand, $L_x$ decreases as $\gamma$ increases, demonstrating that the
null point collapse is inhibited [Fig.~\ref{scaling}(c)]. Even for $\gamma=20$, $L_x$ essentially
measures the current sheet `thickness'.


\section{Relation to analytical solutions}\label{analytical}
\subsection{Dynamic accessibility}
We now investigate the relation between our simulation results and previous analytical solutions for incompressible plasmas.
In the steady-state solutions of Craig {\it et al.}
\cite{craig1995,craig1996} 
the assumption of incompressibility leads to a symmetry between ${\bf B}$ and
${\bf v}$ in the MHD equations. Progress is then made by defining a 3D
current-free `background field', upon which disturbance fields of
low-dimensionality  are super-imposed. This necessarily results in current
sheets which are also of reduced dimensionality.
The solutions are sometimes referred to as `reconnective 
annihilation'\cite{priestetal2000}, since they contain current sheets of infinite extent in at
least one direction, and as a result the plasma advects field lines across
either the spine or the fan, but they only diffuse towards the other of these
(through the current sheet). It might be expected that the infinite nature of
the current sheets is due to the severe analytical restriction of
low-dimensionality `disturbance fields'. However, as we have seen above,
applying shearing boundary motions to the spine footpoints of the null indeed
results in a quasi-planar current sheet in the fan plane, albeit only for
large $\gamma$. 

Of great importance for any steady-state solution is its dynamic
accessibility under a time-dependent evolution. Investigations into the dynamic
accessibility of two-dimensional 
\cite{craig1995} solutions have been carried out by various authors
(e.g.~Refs.~[\onlinecite{watson2004, tassi2005}]). The results of the previous section
provide strong 
evidence that in a fully dynamic and fully 3D (yet incompressible) system, the
fan current sheet solutions are indeed dynamically accessible. One further
question which presents itself here is whether in fact the spine current
solutions are also dynamically accessible. In the analytical solutions, a
tubular spine current results from shearing perturbations of the fan
plane. This is investigated in Section \ref{fandrivesec}.

\subsection{Breakdown of analytical solutions}
It appears that in the incompressible limit, fan current solutions are
dynamically accessible, and (at least qualitatively) provide an accurate snapshot
of the dynamical and fully 3D behaviour. However, in the case of a
compressible plasma this appears not to be the case. In order to understand why
this is, we must examine the force balance which exists in the analytical
solutions. 

The method of the analytical solutions is based upon taking the vector product
(`curl') of the momentum equation, and solving this in conjunction with the
induction equation. The pressure can then be calculated {\it a
  posteriori}. 
However, it has been realised \cite{priest1996b,litvinenko1996,inverarity1996,craig1997}
that this places a limit on the maximum current (or reconnection rate) which
can be attained in these `flux-pile-up' solutions, since the current sheet
must be maintained by a large pressure at infinity. For current values above
some limit, the pressure required is unphysically large.

We can similarly examine the plasma pressure (or pressure gradient) which exists within the current
sheet itself. 
In the steady-state fan current solution of Craig {\it et al.}
\cite{craigetal1995}, the magnetic and velocity fields are defined by
\begin{displaymath}
{\bf B} = \lambda{\bf P} + Y(x) {\bf {\hat y}} + Z(x) {\bf {\hat z}}, \quad
{\bf v} = {\bf P} + \lambda Y(x) {\bf {\hat y}} + \lambda Z(x) {\bf {\hat z}},
\end{displaymath}
\vspace{-0.7cm}
\begin{displaymath}
{\bf P} = \alpha \left( -x, \kappa y, (1-\kappa) z\right).
\end{displaymath}
$\lambda$, $\kappa$, $\alpha$ constant, $0\leq\kappa\leq1$. The pressure is
found from the momentum equation, and the pressure gradient perpendicular to
the fan plane is given by
\begin{displaymath}
\left.\frac{\partial p}{\partial x}\right|_{x=0} = 
-\alpha\lambda \left( \kappa y\frac{\partial Y}{\partial x} +
(1-\kappa)z\frac{\partial Z}{\partial x} \right).
\end{displaymath}
Solving the induction equation for $Y$ and $Z$ (see
Ref.~[\onlinecite{heerikhuisen2004}]) reveals that in the current sheet, $\partial Y /
\partial x \sim \eta^{-(1+\kappa)/2}$, $\partial Z / \partial x \sim
\eta^{-(2-\kappa)/2}$.
Thus in 
the current sheet we require a pressure gradient which scales as a negative
power of $\eta$, which becomes extremely large at realistic values of $\eta$ for
astrophysical plasmas. Note though that the strongest pressure restriction
occurs in the degenerate 2D case ($\kappa=0$ or $\kappa=1$). Once the pressure
gradient can no longer accommodate the 
huge Lorentz force within the sheet, the null point will begin to collapse,
and the strict planar nature of the fan plane and current sheet will be lost  
(note that the Lorentz force always points in the direction which further
closes the angle between spine and fan, while the pressure gradient
acts in the opposite sense). With the symmetry of the system broken, the analytical solutions can no longer describe the behaviour, and we can expect the nature of the current sheet to be significantly altered. 
A similar argument has been made by Ma {\it et
  al.} \cite{ma1995} for the case of disturbances perpendicular to a 2D planar
X-point---they found that once the strict symmetry of the system was broken,
qualitatively very different behaviour resulted.

Leaving the steady-state solutions and examining instead the time-dependent fan current sheet solutions
\cite{craig1998}, one arrives at a similar conclusion. In this case, the
time-dependent pressure gradient force in the $x$-direction in the ideal
localisation phase is given by 
\begin{displaymath}
\frac{\partial p}{\partial x} \sim - e^{\alpha^- (1+\kappa)t}
\end{displaymath}
for one disturbance component ($\alpha^-$ is a constant which determines the relative strengths of the background magnetic and plasma flow fields). This peaks once resistive dissipation becomes important
and the current density reaches a maximum value, when we have 
\begin{displaymath}
\frac{\partial p}{\partial x} \sim \left(\frac{\alpha^-\kappa}{\eta}\right)^
{\frac{1+\kappa}{2}}.
\end{displaymath}
The contribution of the other disturbance component is obtained by replacing
$\kappa$ by $(1-\kappa)$ in each of the above, $0 \leq \kappa \leq 1$. Thus
the plasma pressure 
force in the $x$-direction (or symmetry-breaking direction) increases
exponentially in time, in order to counteract the effect of the increasing
Lorentz force. For sufficiently small $\eta$, the pressure force will no
longer be able to balance the Lorentz force during this localisation process,
and the symmetry of the configuration will be lost.

\begin{figure}
\centering
\includegraphics[scale=0.55]{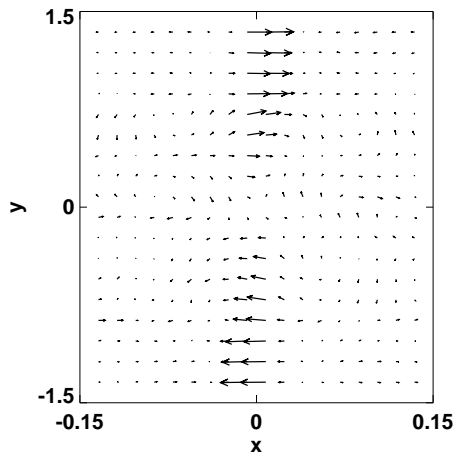}
\caption{Pressure gradient at the time of maximum current in the $z=0$ plane,
  for the run with $\gamma=10$, with driving across the spine.}
\label{gradp}
\end{figure}

The effect of the pressure gradient within our simulations is shown in
Fig.~\ref{gradp}. Here, vectors of $\nabla p$ are plotted in the $z=0$ plane
for $\gamma=10$ at the time of the peak current. It is clear that the
pressure gradient force behaves exactly as described---its effect is
localised primarily within the current sheet (near the $x=0$ plane; compare with Fig.~\ref{jxy}), 
and is directed in such a sense
as to oppose the collapse of the fan surface and current sheet.

The fact that the geometry of the current sheet which we observe in our
compressible simulations is very different to that of the analytical solutions
is not completely unprecedented. In fact, in laboratory experiments examining the
formation of current sheets at 3D nulls, Bogdanov {\it et al}.
\cite{bogdanov1994} made a similar observation. They too found a current
sheet forming at a finite angle to the global directions of both the null spine and fan, which had not been expected from prior self-similar analytical
solutions \cite{bulanov1985}. However, it is interesting to
observe that the incompressible solutions are indeed recovered in the
limit of large $\gamma$, even though we make no assumption regarding the
dimensionality of any fields in the solution.

Note finally that all of the arguments given above carry through to the 2D
case. Thus our results for the 2D null, when compared with the solution of
Ref.~[\onlinecite{craig1995}], can be explained by similar reasoning.


\section{Driving across the fan}\label{fandrivesec}
We now consider the case where the fan of the null is sheared rather than the
spine. We re-run the simulations with $B=B_0(x,-2y,z)$, and again drive in the
$y$-direction on the $x$-boundaries. This time we use a uni-directional
driving profile, which has the disadvantage of compressing the plasma at the
boundaries, causing
a few extra numerical difficulties, but has the advantage of shearing the fan
plane in the same direction over the whole $yz$-plane for each
$x$-boundary. Specifically, we take
\begin{equation}
{\bf v} = V_0(t) \pi \left( 1 - \tanh^2(A_{y}y/Y_l) \right) 
\left( 1 - \tanh^2(A_{z}z/Z_l) \right) {\hat {\bf y}},
\end{equation}
where $V_0$ is again defined by Eq.~(\ref{tdepend}). We take $v_0=0.02$,
$\tau=1.8$, $A_y=12$, $A_z=5$, domain dimensions $X_l=0.5$, $Y_l=Z_l=3$, $\beta=0.05$ and
$B_0=2$ (so that the travel time for the disturbance, which propagates at the
Alfv{\' e}n speed, to reach the null is approximately the same as in the spine shearing
cases).

\begin{figure}
\centering
\includegraphics[scale=0.6]{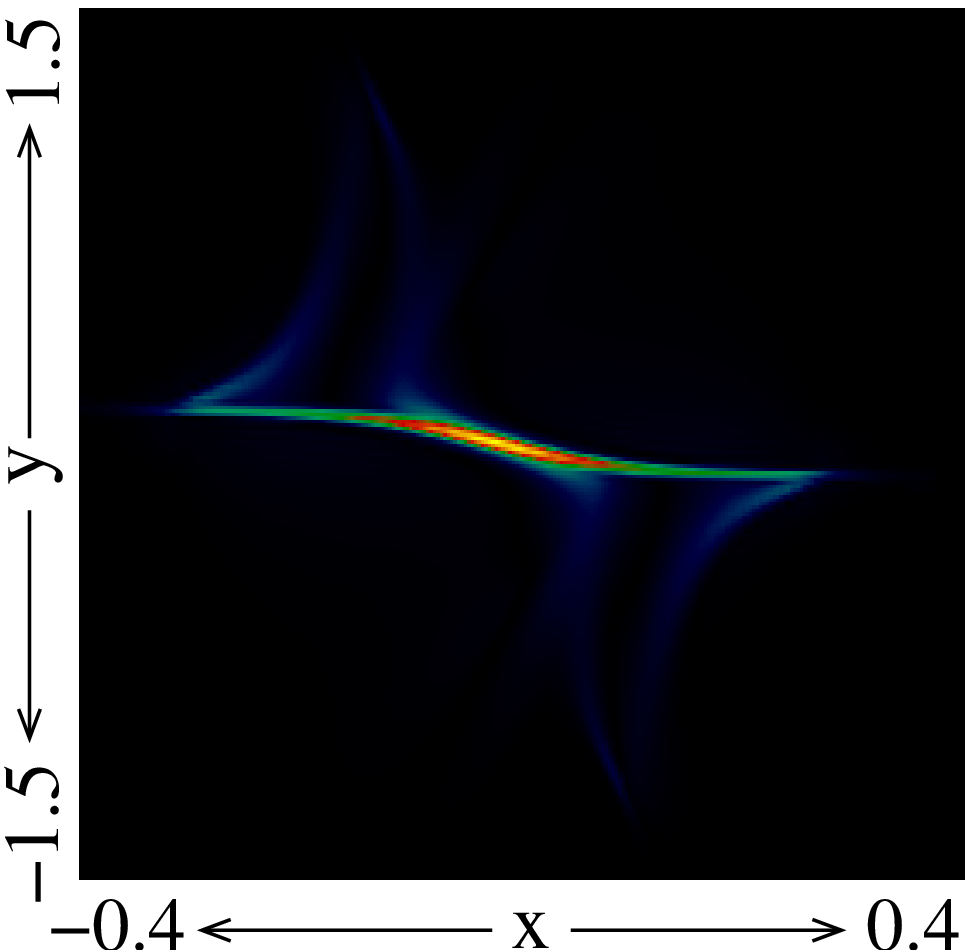}
\includegraphics[scale=0.6]{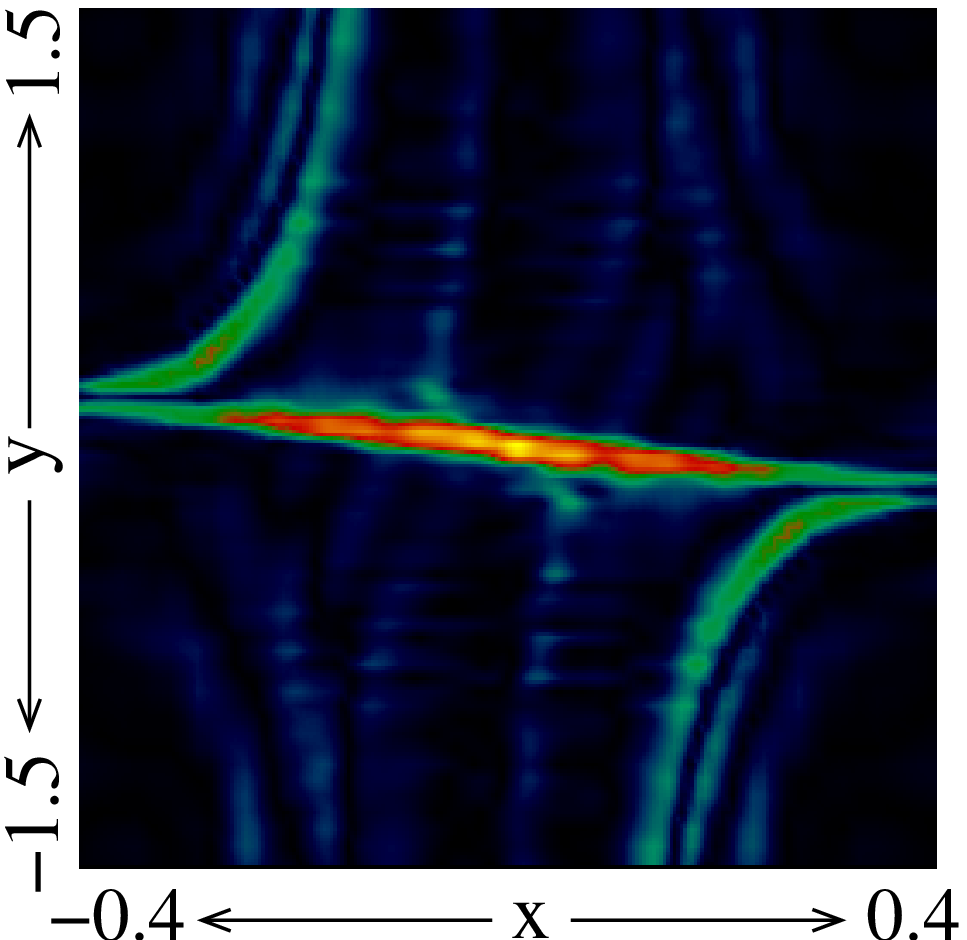}
\caption{(Colour online) Current density $|{\bf J}|$ in the $z=0$ plane, at the time of its temporal
  peak, for (a) $\gamma=5/3$, and (b) $\gamma=10$, for the case of driving across the fan.}
\label{jxyfan}
\end{figure}

The evolution of the null point for an ideal monatomic gas ($\gamma=5/3$) is
very similar to the case where the spine is driven. Once again the disturbance
focuses towards the null point, this time along its fan, and drives it to
collapse. A current sheet 
which spans the spine and fan results [Fig.~\ref{jxyfan}(a)]. This is expected by comparison with
the behaviour of wave-like shear perturbations
\cite{rickard1996,pontingalsgaard2007}. However, 
in the incompressible analytical solution of Craig \& Fabling
\cite{craig1996}, a shear of the fan leads to tubular current structures
aligned to the spine of the null.

Examining the behaviour for larger values of $\gamma$, we find that
compressibility seems to have a similar effect to the spine driving case, but
spine current sheets do not develop. Specifically, decreasing the
compressibility again means that the null does not collapse to the same
extent, though rather than spreading along the spine as predicted by the analytical solutions, the current again
spreads along the fan [Fig.~\ref{jxyfan}(b)].

These results provide strong evidence that spine current sheets are not
dynamically accessible, at least in the absence of strong (super-Alfv{\'
  e}nic) inflows to drive the localisation. This result has previously been
anticipated by Titov {\it et al.} \cite{titov2004}. We instead expect tubular
spine current structures to be associated with rotational motions, see 
Ref.~[\onlinecite{pontingalsgaard2007}]. Within these tubular structures, we expect the current to flow
parallel to the spine, corresponding to field lines spiralling around the spine. By
contrast, the current in the incompressible `spine current' solutions \cite{craig1996} is
directed parallel to the (undisturbed) fan plane (while being localised close
to the spine).


\section{Summary}\label{conc}
We have presented the results of 3D resistive MHD simulations of a driven 3D
null point. We focussed on the effect of moving from a compressible plasma
towards an incompressible one, by varying the ratio of specific heats,
$\gamma$, in our simulations. This was found to strongly affect the resulting
current sheet formation, both qualitatively and quantitatively.

We considered first the case where the spine of the null is sheared from the
boundaries. 
For an ideal, monatomic plasma ($\gamma=5/3$, compressible), the spine and fan
of the null collapse towards one another, and a strongly focused current
sheet forms at the null, locally spanning the spine and fan. However, as
$\gamma$ is increased, the collapse of the null, and in particular of the fan
plane, is suppressed. The current sheet spreads increasingly along the fan
surface, which remains increasingly planar throughout the simulation runs. In
addition, rather than forming a stagnation point flow as the null collapses,
the plasma flow within the domain stays approximately parallel to the planar fan surface
for large $\gamma$. The same effect was found when $\beta$ was
increased rather than $\gamma$, due to the physically similar nature of increasing either parameter, as discussed earlier. 
Quantitatively, the peak current and peak reconnection rate both drop
significantly as $\gamma$ (or $\beta$) is increased (see also Ref.~[\onlinecite{pontincraig2005}]). 

Considering the case where the boundary shearing was applied across the fan plane of the null
rather than the spine, we found similar behaviour. In particular, the
null point collapse is suppressed, and a more spatially diffuse current
structure is found, localised to the fan surface. Our results provide strong
evidence that the steady-state analytical fan current sheet solutions of Craig
{\it et al.} \cite{craig1995} are in fact dynamically accessible in a fully 3D,
incompressible plasma. However, they also imply that the equivalent spine
current sheet solutions \cite{craig1996} are not. Examining the fan current
sheet solutions, it appears that the reason why they break down in a
compressible plasma is the enormous pressure gradients which are required to
maintain the imposed symmetry. These pressure gradients scale inversely with
the resistivity, and so in astrophysical plasmas become unphysically large.

\section{Acknowledgements}
This work was supported by the Department of Energy, Grant
No.~DE-FG02-05ER54832, by the National Science Foundation, Grant
Nos.~ATM-0422764 and ATM-0543202 and by NASA Grant No.~NNX06AC19G. 
K.~G.~was supported by
the Carlsberg Foundation in the form of a fellowship. Computations were
performed on the Zaphod beowulf cluster which was in part funded by the Major
Research Instrumentation program of the National Science Foundation, grant
ATM-0424905. 



\begin{thebibliography}{35}
\expandafter\ifx\csname natexlab\endcsname\relax\def\natexlab#1{#1}\fi
\expandafter\ifx\csname bibnamefont\endcsname\relax
  \def\bibnamefont#1{#1}\fi
\expandafter\ifx\csname bibfnamefont\endcsname\relax
  \def\bibfnamefont#1{#1}\fi
\expandafter\ifx\csname citenamefont\endcsname\relax
  \def\citenamefont#1{#1}\fi
\expandafter\ifx\csname url\endcsname\relax
  \def\url#1{\texttt{#1}}\fi
\expandafter\ifx\csname urlprefix\endcsname\relax\def\urlprefix{URL }\fi
\providecommand{\bibinfo}[2]{#2}
\providecommand{\eprint}[2][]{\url{#2}}

\bibitem[{\citenamefont{Klapper et~al.}(1996)\citenamefont{Klapper, Rado, and
  Tabor}}]{klapper1996}
\bibinfo{author}{\bibfnamefont{I.}~\bibnamefont{Klapper}},
  \bibinfo{author}{\bibfnamefont{A.}~\bibnamefont{Rado}}, \bibnamefont{and}
  \bibinfo{author}{\bibfnamefont{M.}~\bibnamefont{Tabor}},
  \bibinfo{journal}{Phys. Plasmas} \textbf{\bibinfo{volume}{3}},
  \bibinfo{pages}{4281} (\bibinfo{year}{1996}).

\bibitem[{\citenamefont{Priest and Titov}(1996)}]{priest1996}
\bibinfo{author}{\bibfnamefont{E.~R.} \bibnamefont{Priest}} \bibnamefont{and}
  \bibinfo{author}{\bibfnamefont{V.~S.} \bibnamefont{Titov}},
  \bibinfo{journal}{Phil. Trans. R. Soc. Lond. A}
  \textbf{\bibinfo{volume}{354}}, \bibinfo{pages}{2951} (\bibinfo{year}{1996}).

\bibitem[{\citenamefont{Bulanov and Sakai}(1997)}]{bulanov1997}
\bibinfo{author}{\bibfnamefont{S.~V.} \bibnamefont{Bulanov}} \bibnamefont{and}
  \bibinfo{author}{\bibfnamefont{J.}~\bibnamefont{Sakai}}, \bibinfo{journal}{J.
  Phys. Soc. Jpn.} \textbf{\bibinfo{volume}{66}}, \bibinfo{pages}{3477}
  (\bibinfo{year}{1997}).

\bibitem[{\citenamefont{Pontin and Craig}(2005)}]{pontincraig2005}
\bibinfo{author}{\bibfnamefont{D.~I.} \bibnamefont{Pontin}} \bibnamefont{and}
  \bibinfo{author}{\bibfnamefont{I.~J.~D.} \bibnamefont{Craig}},
  \bibinfo{journal}{Phys.~Plasmas} \textbf{\bibinfo{volume}{12}},
  \bibinfo{pages}{072112} (\bibinfo{year}{2005}).

\bibitem[{\citenamefont{Pontin et~al.}(2007)\citenamefont{Pontin,
  Bhattacharjee, and Galsgaard}}]{pontinbhat2007}
\bibinfo{author}{\bibfnamefont{D.~I.} \bibnamefont{Pontin}},
  \bibinfo{author}{\bibfnamefont{A.}~\bibnamefont{Bhattacharjee}},
  \bibnamefont{and} \bibinfo{author}{\bibfnamefont{K.}~\bibnamefont{Galsgaard}}
  (\bibinfo{year}{2007}), \bibinfo{note}{Current sheet formation and non-ideal
  behaviour at 3D magnetic null points, to appear in {\it Phys.~Plasmas}}.

\bibitem[{\citenamefont{Longcope et~al.}(2003)\citenamefont{Longcope, Brown,
  and Priest}}]{longcope2003}
\bibinfo{author}{\bibfnamefont{D.~W.} \bibnamefont{Longcope}},
  \bibinfo{author}{\bibfnamefont{D.~S.} \bibnamefont{Brown}}, \bibnamefont{and}
  \bibinfo{author}{\bibfnamefont{E.~R.} \bibnamefont{Priest}},
  \bibinfo{journal}{Phys. Plasmas} \textbf{\bibinfo{volume}{10}},
  \bibinfo{pages}{3321} (\bibinfo{year}{2003}).

\bibitem[{\citenamefont{Close et~al.}(2005)\citenamefont{Close, Parnell, and
  Priest}}]{close2004}
\bibinfo{author}{\bibfnamefont{R.~M.} \bibnamefont{Close}},
  \bibinfo{author}{\bibfnamefont{C.~E.} \bibnamefont{Parnell}},
  \bibnamefont{and} \bibinfo{author}{\bibfnamefont{E.~R.}
  \bibnamefont{Priest}}, \bibinfo{journal}{Solar Phys.}
  \textbf{\bibinfo{volume}{225}}, \bibinfo{pages}{21} (\bibinfo{year}{2005}).

\bibitem[{\citenamefont{{Fletcher} et~al.}(2001)\citenamefont{{Fletcher},
  {Metcalf}, {Alexander}, {Brown}, and {Ryder}}}]{fletcher2001}
\bibinfo{author}{\bibfnamefont{L.}~\bibnamefont{{Fletcher}}},
  \bibinfo{author}{\bibfnamefont{T.~R.} \bibnamefont{{Metcalf}}},
  \bibinfo{author}{\bibfnamefont{D.}~\bibnamefont{{Alexander}}},
  \bibinfo{author}{\bibfnamefont{D.~S.} \bibnamefont{{Brown}}},
  \bibnamefont{and} \bibinfo{author}{\bibfnamefont{L.~A.}
  \bibnamefont{{Ryder}}}, \bibinfo{journal}{Astrophys. J.}
  \textbf{\bibinfo{volume}{554}}, \bibinfo{pages}{451} (\bibinfo{year}{2001}).

\bibitem[{\citenamefont{Ugarte-Urra et~al.}(2007)\citenamefont{Ugarte-Urra,
  Warren, and Winebarger}}]{ugarteurra2007}
\bibinfo{author}{\bibfnamefont{I.}~\bibnamefont{Ugarte-Urra}},
  \bibinfo{author}{\bibfnamefont{H.~P.} \bibnamefont{Warren}},
  \bibnamefont{and} \bibinfo{author}{\bibfnamefont{A.~R.}
  \bibnamefont{Winebarger}} (\bibinfo{year}{2007}), \bibinfo{note}{The magnetic
  topology of coronal mass ejection sources, {\it Astrophys.~J.}, in press}.

\bibitem[{\citenamefont{Xiao et~al.}(2006)\citenamefont{Xiao, Wang, Pu, Zhao,
  Wang, Ma, Fu, Kivelson, Liu, Zong et~al.}}]{xiao2006}
\bibinfo{author}{\bibfnamefont{C.~J.} \bibnamefont{Xiao}},
  \bibinfo{author}{\bibfnamefont{X.~G.} \bibnamefont{Wang}},
  \bibinfo{author}{\bibfnamefont{Z.~Y.} \bibnamefont{Pu}},
  \bibinfo{author}{\bibfnamefont{H.}~\bibnamefont{Zhao}},
  \bibinfo{author}{\bibfnamefont{J.~X.} \bibnamefont{Wang}},
  \bibinfo{author}{\bibfnamefont{Z.~W.} \bibnamefont{Ma}},
  \bibinfo{author}{\bibfnamefont{S.~Y.} \bibnamefont{Fu}},
  \bibinfo{author}{\bibfnamefont{M.~G.} \bibnamefont{Kivelson}},
  \bibinfo{author}{\bibfnamefont{Z.~X.} \bibnamefont{Liu}},
  \bibinfo{author}{\bibfnamefont{Q.~G.} \bibnamefont{Zong}},
  \bibnamefont{et~al.}, \bibinfo{journal}{Nature Physics}
  \textbf{\bibinfo{volume}{2}}, \bibinfo{pages}{478} (\bibinfo{year}{2006}).

\bibitem[{\citenamefont{Bogdanov et~al.}(1994)\citenamefont{Bogdanov, Burilina,
  Markov, and Frank}}]{bogdanov1994}
\bibinfo{author}{\bibfnamefont{S.~Y.} \bibnamefont{Bogdanov}},
  \bibinfo{author}{\bibfnamefont{V.~B.} \bibnamefont{Burilina}},
  \bibinfo{author}{\bibfnamefont{V.~S.} \bibnamefont{Markov}},
  \bibnamefont{and} \bibinfo{author}{\bibfnamefont{A.~G.} \bibnamefont{Frank}},
  \bibinfo{journal}{JETP Lett.} \textbf{\bibinfo{volume}{59}},
  \bibinfo{pages}{537} (\bibinfo{year}{1994}).

\bibitem[{\citenamefont{Parnell et~al.}(1996)\citenamefont{Parnell, Smith,
  Neukirch, and Priest}}]{parnell1996}
\bibinfo{author}{\bibfnamefont{C.~E.} \bibnamefont{Parnell}},
  \bibinfo{author}{\bibfnamefont{J.~M.} \bibnamefont{Smith}},
  \bibinfo{author}{\bibfnamefont{T.}~\bibnamefont{Neukirch}}, \bibnamefont{and}
  \bibinfo{author}{\bibfnamefont{E.~R.} \bibnamefont{Priest}},
  \bibinfo{journal}{Phys.~Plasmas} \textbf{\bibinfo{volume}{3}},
  \bibinfo{pages}{759} (\bibinfo{year}{1996}).

\bibitem[{\citenamefont{Lau and Finn}(1990)}]{lau1990}
\bibinfo{author}{\bibfnamefont{Y.~T.} \bibnamefont{Lau}} \bibnamefont{and}
  \bibinfo{author}{\bibfnamefont{J.~M.} \bibnamefont{Finn}},
  \bibinfo{journal}{Astrophys. J.} \textbf{\bibinfo{volume}{350}},
  \bibinfo{pages}{672} (\bibinfo{year}{1990}).

\bibitem[{\citenamefont{Pontin et~al.}(2004)\citenamefont{Pontin, Hornig, and
  Priest}}]{pontin2004}
\bibinfo{author}{\bibfnamefont{D.~I.} \bibnamefont{Pontin}},
  \bibinfo{author}{\bibfnamefont{G.}~\bibnamefont{Hornig}}, \bibnamefont{and}
  \bibinfo{author}{\bibfnamefont{E.~R.} \bibnamefont{Priest}},
  \bibinfo{journal}{Geophys. Astrophys. Fluid Dynamics}
  \textbf{\bibinfo{volume}{98}}, \bibinfo{pages}{407} (\bibinfo{year}{2004}).

\bibitem[{\citenamefont{Pontin et~al.}(2005)\citenamefont{Pontin, Hornig, and
  Priest}}]{pontinhornig2005}
\bibinfo{author}{\bibfnamefont{D.~I.} \bibnamefont{Pontin}},
  \bibinfo{author}{\bibfnamefont{G.}~\bibnamefont{Hornig}}, \bibnamefont{and}
  \bibinfo{author}{\bibfnamefont{E.~R.} \bibnamefont{Priest}},
  \bibinfo{journal}{Geophys. Astrophys. Fluid Dynamics}
  \textbf{\bibinfo{volume}{99}}, \bibinfo{pages}{77} (\bibinfo{year}{2005}).

\bibitem[{\citenamefont{Craig and Henton}(1995)}]{craig1995}
\bibinfo{author}{\bibfnamefont{I.~J.~D.} \bibnamefont{Craig}} \bibnamefont{and}
  \bibinfo{author}{\bibfnamefont{S.~M.} \bibnamefont{Henton}},
  \bibinfo{journal}{Astrophys. J.} \textbf{\bibinfo{volume}{450}},
  \bibinfo{pages}{280} (\bibinfo{year}{1995}).

\bibitem[{\citenamefont{Craig and Fabling}(1996)}]{craig1996}
\bibinfo{author}{\bibfnamefont{I.~J.~D.} \bibnamefont{Craig}} \bibnamefont{and}
  \bibinfo{author}{\bibfnamefont{R.~B.} \bibnamefont{Fabling}},
  \bibinfo{journal}{Astrophys. J.} \textbf{\bibinfo{volume}{462}},
  \bibinfo{pages}{969} (\bibinfo{year}{1996}).

\bibitem[{\citenamefont{Craig and Fabling}(1998)}]{craig1998}
\bibinfo{author}{\bibfnamefont{I.~J.~D.} \bibnamefont{Craig}} \bibnamefont{and}
  \bibinfo{author}{\bibfnamefont{R.~B.} \bibnamefont{Fabling}},
  \bibinfo{journal}{Phys. Plasmas} \textbf{\bibinfo{volume}{5}},
  \bibinfo{pages}{635} (\bibinfo{year}{1998}).

\bibitem[{\citenamefont{Galsgaard and Nordlund}(1997)}]{galsgaard1997}
\bibinfo{author}{\bibfnamefont{K.}~\bibnamefont{Galsgaard}} \bibnamefont{and}
  \bibinfo{author}{\bibfnamefont{A.}~\bibnamefont{Nordlund}},
  \bibinfo{journal}{J. Geophys. Res.} \textbf{\bibinfo{volume}{102}},
  \bibinfo{pages}{231} (\bibinfo{year}{1997}).

\bibitem[{\citenamefont{Archontis et~al.}(2004)\citenamefont{Archontis,
  Moreno-Insertis, Galsgaard, Hood, and O'Shea}}]{archontis2004}
\bibinfo{author}{\bibfnamefont{V.}~\bibnamefont{Archontis}},
  \bibinfo{author}{\bibfnamefont{F.}~\bibnamefont{Moreno-Insertis}},
  \bibinfo{author}{\bibfnamefont{K.}~\bibnamefont{Galsgaard}},
  \bibinfo{author}{\bibfnamefont{A.}~\bibnamefont{Hood}}, \bibnamefont{and}
  \bibinfo{author}{\bibfnamefont{E.}~\bibnamefont{O'Shea}},
  \bibinfo{journal}{Astron.~Astrophys.} \textbf{\bibinfo{volume}{426}},
  \bibinfo{pages}{1074} (\bibinfo{year}{2004}).

\bibitem[{\citenamefont{Priest and Forbes}(2000)}]{priest2000}
\bibinfo{author}{\bibfnamefont{E.~R.} \bibnamefont{Priest}} \bibnamefont{and}
  \bibinfo{author}{\bibfnamefont{T.~G.} \bibnamefont{Forbes}},
  \emph{\bibinfo{title}{Magnetic reconnection: MHD theory and applications}}
  (\bibinfo{publisher}{Cambridge University Press, Cambridge},
  \bibinfo{year}{2000}).

\bibitem[{\citenamefont{Priest et~al.}(2000)\citenamefont{Priest, Titov,
  Grundy, and Hood}}]{priestetal2000}
\bibinfo{author}{\bibfnamefont{E.~R.} \bibnamefont{Priest}},
  \bibinfo{author}{\bibfnamefont{V.~S.} \bibnamefont{Titov}},
  \bibinfo{author}{\bibfnamefont{R.~E.~G.} \bibnamefont{Grundy}},
  \bibnamefont{and} \bibinfo{author}{\bibfnamefont{A.~W.} \bibnamefont{Hood}},
  \bibinfo{journal}{Proc.~Roy.~Soc.~Lond.~A} \textbf{\bibinfo{volume}{456}},
  \bibinfo{pages}{1821} (\bibinfo{year}{2000}).

\bibitem[{\citenamefont{Watson and Porcelli}(2004)}]{watson2004}
\bibinfo{author}{\bibfnamefont{P.~G.} \bibnamefont{Watson}} \bibnamefont{and}
  \bibinfo{author}{\bibfnamefont{F.}~\bibnamefont{Porcelli}},
  \bibinfo{journal}{Astrophys. J.} \textbf{\bibinfo{volume}{617}},
  \bibinfo{pages}{1353} (\bibinfo{year}{2004}).

\bibitem[{\citenamefont{Tassi et~al.}(2005)\citenamefont{Tassi, Titov, and
  Hornig}}]{tassi2005}
\bibinfo{author}{\bibfnamefont{E.}~\bibnamefont{Tassi}},
  \bibinfo{author}{\bibfnamefont{V.~S.} \bibnamefont{Titov}}, \bibnamefont{and}
  \bibinfo{author}{\bibfnamefont{G.}~\bibnamefont{Hornig}},
  \bibinfo{journal}{Phys. Plasmas} \textbf{\bibinfo{volume}{12}},
  \bibinfo{pages}{112902} (\bibinfo{year}{2005}).

\bibitem[{\citenamefont{Priest}(1996)}]{priest1996b}
\bibinfo{author}{\bibfnamefont{E.~R.} \bibnamefont{Priest}},
  \emph{\bibinfo{title}{in Solar and astrophysical MHD flows}}
  (\bibinfo{publisher}{Kluwer}, \bibinfo{year}{1996}), pp.
  \bibinfo{pages}{151--170}.

\bibitem[{\citenamefont{Litvinenko et~al.}(1996)\citenamefont{Litvinenko,
  Forbes, and Priest}}]{litvinenko1996}
\bibinfo{author}{\bibfnamefont{Y.~E.} \bibnamefont{Litvinenko}},
  \bibinfo{author}{\bibfnamefont{T.~G.} \bibnamefont{Forbes}},
  \bibnamefont{and} \bibinfo{author}{\bibfnamefont{E.~R.}
  \bibnamefont{Priest}}, \bibinfo{journal}{Solar Phys.}
  \textbf{\bibinfo{volume}{167}}, \bibinfo{pages}{445} (\bibinfo{year}{1996}).

\bibitem[{\citenamefont{Inverarity and Priest}(1996)}]{inverarity1996}
\bibinfo{author}{\bibfnamefont{G.~W.} \bibnamefont{Inverarity}}
  \bibnamefont{and} \bibinfo{author}{\bibfnamefont{E.~R.}
  \bibnamefont{Priest}}, \bibinfo{journal}{Phys. Plasmas}
  \textbf{\bibinfo{volume}{3}}, \bibinfo{pages}{3591} (\bibinfo{year}{1996}).

\bibitem[{\citenamefont{Craig et~al.}(1997)\citenamefont{Craig, Fabling, and
  Watson}}]{craig1997}
\bibinfo{author}{\bibfnamefont{I.~J.~D.} \bibnamefont{Craig}},
  \bibinfo{author}{\bibfnamefont{R.~B.} \bibnamefont{Fabling}},
  \bibnamefont{and} \bibinfo{author}{\bibfnamefont{P.~G.}
  \bibnamefont{Watson}}, \bibinfo{journal}{Astrophys. J.}
  \textbf{\bibinfo{volume}{485}}, \bibinfo{pages}{383} (\bibinfo{year}{1997}).

\bibitem[{\citenamefont{Craig et~al.}(1995)\citenamefont{Craig, Fabling,
  Henton, and Rickard}}]{craigetal1995}
\bibinfo{author}{\bibfnamefont{I.~J.~D.} \bibnamefont{Craig}},
  \bibinfo{author}{\bibfnamefont{R.~B.} \bibnamefont{Fabling}},
  \bibinfo{author}{\bibfnamefont{S.~M.} \bibnamefont{Henton}},
  \bibnamefont{and} \bibinfo{author}{\bibfnamefont{G.~J.}
  \bibnamefont{Rickard}}, \bibinfo{journal}{Astrophys. J. Lett.}
  \textbf{\bibinfo{volume}{455}}, \bibinfo{pages}{L197} (\bibinfo{year}{1995}).

\bibitem[{\citenamefont{Heerikhuisen and Craig}(2004)}]{heerikhuisen2004}
\bibinfo{author}{\bibfnamefont{J.}~\bibnamefont{Heerikhuisen}}
  \bibnamefont{and} \bibinfo{author}{\bibfnamefont{I.~J.~D.}
  \bibnamefont{Craig}}, \bibinfo{journal}{Solar Phys.}
  \textbf{\bibinfo{volume}{222}}, \bibinfo{pages}{95} (\bibinfo{year}{2004}).

\bibitem[{\citenamefont{Ma et~al.}(1995)\citenamefont{Ma, Ng, Wang, and
  Bhattacharjee}}]{ma1995}
\bibinfo{author}{\bibfnamefont{Z.~W.} \bibnamefont{Ma}},
  \bibinfo{author}{\bibfnamefont{C.~S.} \bibnamefont{Ng}},
  \bibinfo{author}{\bibfnamefont{X.}~\bibnamefont{Wang}}, \bibnamefont{and}
  \bibinfo{author}{\bibfnamefont{A.}~\bibnamefont{Bhattacharjee}},
  \bibinfo{journal}{Phys. Plasmas} \textbf{\bibinfo{volume}{2}},
  \bibinfo{pages}{3184} (\bibinfo{year}{1995}).

\bibitem[{\citenamefont{Bulanov and Olshanetsky}(1985)}]{bulanov1985}
\bibinfo{author}{\bibfnamefont{S.~V.} \bibnamefont{Bulanov}} \bibnamefont{and}
  \bibinfo{author}{\bibfnamefont{M.~A.} \bibnamefont{Olshanetsky}},
  \bibinfo{journal}{Sov. J. Plasma Phys.} \textbf{\bibinfo{volume}{11}},
  \bibinfo{pages}{425} (\bibinfo{year}{1985}).

\bibitem[{\citenamefont{Rickard and Titov}(1996)}]{rickard1996}
\bibinfo{author}{\bibfnamefont{G.~J.} \bibnamefont{Rickard}} \bibnamefont{and}
  \bibinfo{author}{\bibfnamefont{V.~S.} \bibnamefont{Titov}},
  \bibinfo{journal}{Astrophys. J.} \textbf{\bibinfo{volume}{472}},
  \bibinfo{pages}{840} (\bibinfo{year}{1996}).

\bibitem[{\citenamefont{Pontin and Galsgaard}(2007)}]{pontingalsgaard2007}
\bibinfo{author}{\bibfnamefont{D.~I.} \bibnamefont{Pontin}} \bibnamefont{and}
  \bibinfo{author}{\bibfnamefont{K.}~\bibnamefont{Galsgaard}},
  \bibinfo{journal}{J.~Geophys.~Res.} \textbf{\bibinfo{volume}{112}},
  \bibinfo{pages}{A03103} (\bibinfo{year}{2007}).

\bibitem[{\citenamefont{Titov et~al.}(2004)\citenamefont{Titov, Tassi, and
  Hornig}}]{titov2004}
\bibinfo{author}{\bibfnamefont{V.~S.} \bibnamefont{Titov}},
  \bibinfo{author}{\bibfnamefont{E.}~\bibnamefont{Tassi}}, \bibnamefont{and}
  \bibinfo{author}{\bibfnamefont{G.}~\bibnamefont{Hornig}},
  \bibinfo{journal}{Phys. Plasmas} \textbf{\bibinfo{volume}{11}},
  \bibinfo{pages}{4662} (\bibinfo{year}{2004}).

\end{thebibliography}

\end{document}